\documentstyle[12pt]{article}

\begin{document}

\newcommand{\nc}{\newcommand}

\nc{\beq}{\begin{equation}}
\nc{\eeq}{\end{equation}}
\nc{\beqa}{\begin{eqnarray}} \nc{\eeqa}{\end{eqnarray}}
\nc{\lsim}{\begin{array}{c}\,\sim\vspace{-21pt}\\< \end{array}}
\nc{\gsim}{\begin{array}{c}\sim\vspace{-21pt}\\> \end{array}}
\nc{\el}{{\cal L}}
\nc{\D}{{\cal D}}

\renewcommand{\thefootnote}{\fnsymbol{footnote}}
\setcounter{footnote}{0}
\begin{titlepage}

\def\thefootnote{\fnsymbol{footnote}}

\begin{center}

\hfill YCTP-P26-99\\
\hfill hep-th/9909199\\
\hfill September, 1999\\

\vskip .5in

{\Large \bf Warp Factors and Extended Sources

\smallskip

 in Two Transverse Dimensions }

\vskip .45in

{\large Alan Chodos and Erich Poppitz }

\vskip 0.2in

 {\tt alan.chodos@yale.edu, erich.poppitz@yale.edu}

\vskip 0.2in
{\em
  Department of Physics\\
  Yale University\\
  New Haven\\
   CT 06520-8120, USA}

\end{center}

\vskip .4in

\begin{abstract}

We study the  solutions of the  Einstein equations in
$(d+2)$-dimensions, describing parallel $p$-branes ($p=d-1$) in a
space with two transverse dimensions of positive gaussian
curvature. These solutions generalize the solutions of Deser and
Jackiw of point particle sources in $(2+1)$-dimensional gravity
with cosmological constant. Determination of the metric is reduced
to finding the roots of a simple algebraic equation. These roots
also determine the nontrivial ``warp factors" of the metric at the
positions of the branes. We discuss the possible role of these
solutions and   the importance of ``warp factors" in the context of
the large extra dimensions scenario.

\end{abstract}
\end{titlepage}

\renewcommand{\thepage}{\arabic{page}}
\setcounter{page}{1}
\renewcommand{\thefootnote}{\#\arabic{footnote}}
\setcounter{footnote}{0}

\section{Introduction and summary.}

We study  solutions of the Einstein equations of $d + 2$
dimensional gravity with a cosmological constant. The solutions
represent $(d-1)$-branes embedded in a space with two compact
transverse dimensions with a positive curvature. They generalize
the static solutions of Deser and Jackiw \cite{DJ}, describing
point particles in $2+1$ dimensional gravity with cosmological
constant, to the case of extended objects.

The solutions we find obey Einstein's equations, as well as the
equation of motion of the branes (the geodesic equation).  It is
remarkable that the metric, in appropriate coordinates, can be
found without explicitly solving the Einstein equations. Analyzing
the properties of the solutions is reduced to finding the roots of
a simple algebraic equation.

{}From the point of view of applications to the recently proposed
\cite{ADD}, \cite{extra1} large extra dimensions scenario, the most
important property of our solutions is that they are characterized
by nontrivial ``warp factors" that affect the values of the
parameters of the world-volume theories \cite{RS}. In particular,
the usual relation between the four dimensional Planck scale, the 6
dimensional ``fundamental" scale of gravity, and the volume of the
compactified space is affected by the presence of the warp factors.
This difference can be important if the warp factors significantly
deviate from unity (which, in the present framework can only be
achieved by fine tuning various parameters) (for related papers
with one transverse dimension, see \cite{NK}).

While our recipe for finding solutions is rather general, motivated
by the large extra dimensions proposal, we discuss in some detail
 the solution that describes two 3-branes embedded in six
dimensional space time with negative\footnote{We note that our
conventions for the signs of the curvature are those of
\cite{weinberg}; in addition $\Lambda < 0$  in
(\ref{gravityaction}) corresponds to de Sitter space.} (de Sitter)
cosmological constant. The two branes are located on the north and
south poles of a ``sphere", with a ``wedge" cut out, due to the
deficit angle characteristic of pointlike sources in 2 dimensions.
An important property of the solution is that the metric on the
branes is not flat, but rather de Sitter. While the cosmological
constant
 and the strength of gravity on the ``visible" brane  can be tuned to
 satisfy the experimental constraints, within our ansatz
 there is no solution with flat world volume metric.

This paper is structured as follows. In Section 2, we present the
ansatz and derive the equations of motion. We show that they reduce
to a simple differential equation in one real variable, similar to
the case considered by Deser and Jackiw \cite{DJ}.  In Section 3,
we give a general analysis of the solutions. We show that the form
of the metric (in appropriate coordinates) as well as the
interpretation of its singularities can be obtained without
explicitly solving the differential equation. All information on
the solution can be deduced by finding the appropriate roots of an
algebraic equation. In particular,  these roots determine the
values of the warp factors on the branes. We derive expressions for
the strength of the gravitational coupling, the cosmological
constant on the various branes, and the size of the transverse
space. Finally, in Section 4, we discuss an explicit example, with
$d=4$, which is of interest to the large extra dimension scenario.
We show that the parameters allow fine tuning of the Planck scale
and cosmological constant to values that do not contradict the
observed ones.

\section{The ansatz and equations of motion.}

 The action we consider is that of gravity in $D = d + 2$ dimensions,
 with the usual Einstein  action with cosmological term included:
\beq
\label{gravityaction}
S_{gravity} = - M^{D-2} \int d^D y \sqrt{g} \left( R - 2 \Lambda
\right)~.
\eeq
The indices $M, N = 1,...,D$, while $\mu,\nu = 0,...,d = D - 2$ and
$i,j
= 1,2$ span the rest of the space; the metric has signature
$(-,+,\ldots, +)$. The vacuum Einstein equations follow from the
variation of the gravity action:
\beq
\label{vacuumeinstein}
\delta S_{gravity}= - M^{D-2} \int d^D y \sqrt{g}
\left( - R^{MN} + {1\over 2}  g^{MN}  R - g^{MN} \Lambda \right) \delta g_{MN}
 = 0~.
\eeq
The ``matter" is in the form of $d-1$ dimensional branes, whose
action is proportional to the area of the world surface   they sweep in
the $d + 2$ dimensional spacetime:
\beq
\label{matteraction}
S_{matter} = - \sum_a f_a^d \int d^D y \int d^d
\sigma ~\delta^D(y - X_a(\sigma))~ \sqrt{\tilde{g}},
\eeq
where  $X_a^M (\sigma)$ is the embedding of the world surface in
space time ($a$ runs over the various branes) and
 $\tilde{g}_{\mu \nu} = X^M_{,\mu}~ X^N_{,\nu} ~g_{MN}(y)$
the induced metric and $\tilde{g}\equiv -
\det
\tilde{g}_{\mu
\nu}$ (hereafter we will omit the sum over the various sources,
it will be implicit in all our formulae).
The matter energy
momentum tensor is defined by:
\beqa
\label{matteractionvariation}
\delta S_{matter} &=& {1 \over 2} \int d^D y ~\sqrt{g} ~
T^{MN}(y) \delta g_{MN}(y)\nonumber \\ &=& - {1 \over 2}~f^d~ \int
d^D y
\int d^d \sigma ~\delta^D (y - X(\sigma))~
 \tilde{g}^{1/2}~ X^M_{,\mu} ~X^N_{,\nu} ~\tilde{g}^{\mu
\nu}~
\delta g_{MN}(y)~,
\eeqa
hence
\beq
\label{matterenergymomentum}
\sqrt{g} T^{MN}(y) = - f^d \int d^d \sigma ~\delta^D(y - X(\sigma))~
\tilde{g}^{1/2} ~X^M_{,\mu} ~X^N_{,\nu}~ \tilde{g}^{\mu
\nu}~.
\eeq
The equation of motion of the branes (obtained by varying the
action with respect to the brane coordinates  $X(\sigma)$) is the
``geodesic" equation:
\beq
\label{geodesic}
\nabla^2 X^M ~+ ~
\Gamma^M_{KL} ~\tilde{g}^{\mu \nu}~ X^K_{,\mu}
~X^L_{,\nu} = 0~,
\eeq
where $\nabla^2 = \tilde{g}^{-1/2} \partial_\mu \tilde{g}^{1/2}
\tilde{g}^{\mu \nu} \partial_\nu$ is the covariant D'Alembertian
in the induced metric.

 Our ansatz for the metric includes a nontrivial warp factor $e^{2 \lambda}$,
 depending
 only on the coordinates transverse to the branes (denoted by $x^i$,
$i = 1,2$):
\beq
\label{metricansatz}
d s^2 = e^{2 \lambda (x^i)} \left[ g_{\mu \nu} (x^{\lambda})  d
x^\mu d x^\nu + e^{2 K (x^i)} d x^i d x^i \right]~.
\eeq
We assume that the d-dimensional metric $g_{\mu
\nu}(x^\lambda)$ obeys the equation
\beq
\label{4dmetric}
R_{\mu \nu} - {1\over 2} g_{\mu \nu} R + \alpha g_{\mu \nu} = 0~,
\eeq
so that $\alpha$ becomes a parameter of our solution.

The idea for finding a solution with delta-function sources is  to
first solve the source-free equations of motion. Then, admitting
delta function like curvature singularities of the metric will
allow us to include the effect of the sources. Following this
strategy, in the remainder of this section we   consider the source
free equations.

The Einstein equation following from the variation of the action
with respect to the $\mu, \nu$ components of the metric is:
\beq
\label{munuequation}
d \lambda_{, ii} + K_{, ii} + {d (d-1) \over 2} \lambda_{,i}
\lambda_{,i} = e^{2 \lambda + 2 K} \Lambda - e^{2 K} \alpha ~
\eeq
This equation, away from the singularities, is a consequence of the
other equations of motion. Static sources will lead to
delta-function singularities on the r.h.s. of equation
(\ref{munuequation}). These will appear, as we will show in Section
3, because of singularites of $K_{,ii}$.

The (trace of) the variation of the action w.r.t. the $i,j$
components of the metric gives the following equation:
\beq
\label{singleeqn}
\partial \bar{\partial} \lambda + d ~\partial \lambda ~\bar{\partial} \lambda =
{2 \over d} e^{2 \lambda + 2 K} \Lambda - {2 \over d -2} e^{2K} \alpha ~.
\eeq
We have introduced complex coordinates, $z
= (x^1 + i x^2)/2, \bar{z} = (x^1 - i x^2)/2$, and defined
$\partial=
\partial_{x_1} - i \partial_{x_2}$,
$ \bar{\partial} = \partial_{x_1} + i \partial_{x_2}$ such that
$\partial z = 1, \partial \bar{z} = 0$. On the other hand, the
traceless part of the equations of motion that follow from varying
the action with respect to the $i,j$ components of the metric are
nothing but the Cauchy-Riemann conditions for the function
\beq
\label{V}
V = e^{- 2K - \lambda} \bar{\partial} \lambda~.
\eeq
The function $V$, therefore is holomorphic, $\bar{\partial}V=0$.
This fact will play a central role in our ability to find explicit
solutions.\footnote{Before continuing with the analysis of
solutions with nontrivial warp factors, we note that the
source-free equations of motion (\ref{singleeqn},\ref{V}) admit a
simple solution  with $K
= -  \log ( 1 + 4 |z|^2/\rho^2)$, and $\lambda
= const.$, provided the values of $\Lambda$ and $\alpha$ are
related, so that the r.h.s. of (\ref{singleeqn}) vanishes. These
solutions are of the form $dS_{d-2} \times S^2$, with the radius of
$S^2$, $\rho$, and the cosmological constant of the $dS_{d-2}$,
$\alpha$, related by (\ref{singleeqn}): $\alpha = - 2 \rho^{-2}
(d-2) $, $\Lambda = - 2 \rho^{-2} d $ (for $\lambda = 0$). These
vacuum solutions of the $d+2$ dimensional Einstein equations with
cosmological constant generalize the Nariai solutions \cite{Nariai}
to $d+2$ dimensions. We thank N. Kaloper for pointing out to us the
existence of ref.~\cite{Nariai}.}

We can now further simplify our equations, following \cite{DJ}.
Multiplying (\ref{singleeqn}) by $V$ given by (\ref{V}) and using
$\bar{\partial} V = 0$, we obtain after integrating over $\bar{z}$,
\beq
\label{finaleqn1}
V \partial e^{d \lambda} = {2 \Lambda \over d + 1} e^{ (d + 1)
\lambda} - {2 d \over (d-2) (d -1)} \alpha e^{(d-1) \lambda} + \epsilon(z)~,
\eeq
where $\epsilon(z)$ is an integration constant. Upon introducing
the new complex variable
\beq
\label{xi}
\xi =  \int^z { d w \over V(w)}
\eeq
as well as denoting $e^\lambda = N$, we can rewrite the equation
as:
\beq
\label{finaleqn2}
{\partial \over \partial \xi} N^d = {2 \Lambda \over d + 1} N^{d +
1} - {2 d \over (d-2) (d -1)} \alpha N^{d-1} + \epsilon~.
\eeq
Now upon taking the complex conjugate equation, we see that  the
reality condition for $N$ implies that   $N$ is a function of the
real part of $\xi$ only  and that $\epsilon$ must be a real
constant. Finally,  introduce the variable
\beq
\label{t}
 t ~=~{2 \Lambda \over (d + 1) d}~ \left( \xi + \xi^* \right) ~
 =~{2 \Lambda \over (d + 1) d}~ \left( \int^z { d w \over V(w)} ~+~{\rm h.c.} \right)
\eeq
      and write  the  equation   in terms of the single real variable $t$ (\ref{t}):
\beq
\label{Finaleqn}
f(N) ~d N ~\equiv ~ {N^{d -1} \over P(N) } ~d N ~= d t ~,
\eeq
where $f(N)$ is defined by the first equality, and  the polynomial
$P(N)$ is:
\beq
\label{polynomial}
P(N) ~=~ N^{d+1} - a  N^{d-1} +  b~,
\eeq
with the coefficients $a,b$ defined as follows:
\beq
\label{ab}
a ~=~ {\alpha \over \Lambda} ~{ d (d+1) \over (d-2) (d-1)}~,~~~ b
~=~ {\epsilon \over \Lambda} ~ {d + 1 \over 2}~.
\eeq

This completes our discussion of the equations of motion. We saw
that the equation of motion for the ansatz (\ref{metricansatz})
reduced to a simple differential equation in one real variable
(\ref{Finaleqn}); a general solution can be expressed in terms of
the general solution of this equation  and a holomorphic function
$V$ (\ref{V}). We also note that this provides also a solution to
eq.~(\ref{munuequation})---it is easy to see that (\ref{singleeqn})
together with the holomorphicity of $V$ implies
(\ref{munuequation}), away from possible singularities (which, as
we will see in the next section, appear in $K_{,ii}$).

In the following section we   present a general analysis of the
solutions of (\ref{Finaleqn}).

\section{General analysis of the solutions.}

In this section, we discuss the properties of the general solution
of the equation (\ref{Finaleqn}). To begin, we need to specify the
range of the variable $t$ in (\ref{Finaleqn}). In the cases of
interest to us  $t$ will span the
 entire real axis---consider e.g. the case
  $V(z) = z/c$, so that $t \sim \log |z|$.
  In order to solve (\ref{Finaleqn})
 we need to find a function $F(N)$ such that $F^\prime (N) = f(N)$. Moreover,
  we should  be able to invert $F(N)$ for all values of $t$---then
  $N(t) = F^{-1}(t)$ will satisfy  (\ref{Finaleqn}).

  Even though the equation is rather complicated and the explicit form of $F(N)$ is
  generally unknown, we
  will see that  the quantities of interest of us do not require an explicit
  knowledge of the solution. In fact, the metric can, in appropriate
  coordinates, be expressed in terms of the polynomial $P(N)$, while the ranges of the
  variables are determined by appropriately chosen roots of the polynomial $P(N)$.

  The basic requirement is that we find a function $F(N)$
  which  monotonically changes from $-\infty$ to $\infty$ over some finite range
  of values of $N$. For that to be the case, the function $f(N)$ has to have
  the properties
  (recall that $F^\prime = f$) that between two values of $N$  {\it a}) it does not
  change sign and {\it b}) its modulus approaches infinity at the two boundary
  values of $N$. Consider two ``nearest neighbor" roots of the polynomial $P(N)$, $N_1$ and $N_2$.
  Assume that both $N_1$ and $N_2$ have the same sign.\footnote{This requirement
can be dropped for $d =1$.}
  Now, at the roots of $P(N)$   $f(N)$ blows up; moreover,
 if $N_1$ and $N_2$  have the  same  sign,  $f(N)$  does not
  change sign as $N$ varies between $N_1$ and $N_2$ (we are mostly interested in the case of even $d$).
  Then, between these two roots  $F(N)$ is monotonic and
   approaches $\infty$ at one of the roots, say $N_1$, and $-\infty$
   at the other root, $N_2$.
 Hence, we can invert the equation (\ref{Finaleqn}) and find the function $N(t) = F^{-1}(t)$
 for all values of $t$ on the real axis; moreover, the function $N(t)$ approaches finite values---the roots of
 $P(N)$, $N_1$ and  $N_2$---as $|t| \rightarrow \infty$.

Thus the problem of finding a solution of (\ref{Finaleqn}) is
reduced to finding the condition that there are two real
``nearest-neighbor" roots of $P(N)$ of the same sign. It is easily
seen, e.g. for $d=4$, that this requires that $a > 0$, i.e.
$\alpha$ and $\Lambda$ have the same sign.

We assume now that the parameters $a, b$ (\ref{ab}) are such that
two such roots, $N_1$ and $N_2$ are found (assume for simplicity
that they are both positive). Let us now analyze the properties of
the solution. The metric has the form (\ref{metricansatz}); having
found the function $N(t)$, we can now determine the $N^2 e^{2K}$
factor in the metric by using (\ref{V}) and the equation
(\ref{Finaleqn}):
\beq
\label{exp2K}
N^2 e^{2K} ~=~ {1 \over V} ~\bar{\partial} N ~=~{2 \Lambda \over d
(d + 1) } ~{1 \over |V|^2}~ {P(N) \over N^{d -1}}~.
\eeq
Now note that using
\beqa
\label{ttheta}
d t ~&=&~ { 2 \Lambda \over d (d +1) } \left[ {d z \over V(z)} +
 { d {\bar z} \over {\bar V}({\bar z})} \right] \\
d \theta ~&\equiv&~ i \beta  \left[ {d z\over V(z)}
- {d {\bar z} \over {\bar V}({ \bar z } ) }\right] ~\nonumber
\eeqa
we can rewrite the metric (\ref{metricansatz})  as:
\beqa
\label{metric1}
 d s^2 ~&=&~ N^2 ds_{(d)}^2 ~+~ 4 N^2 e^{2K} d z d \bar{z} \\
 \nonumber \\
\label{metric11}
&=& N^2 ds_{(d)}^2~ +~{d (d+1) \over 2 \Lambda}~{P(N) \over N^{d
-1}}~d t^2 ~+~{ 2 \Lambda \over d (d + 1)} ~{P(N) \over N^{d
-1}}~{d \theta^2 \over \beta^2}~.
\eeqa
Finally, we can change variables from $t$ to $N$ using
(\ref{Finaleqn}) and write the metric as:
\beq
\label{metric2}
d s^2 ~=~ N^2 ds_{(d)}^2 ~+~ {d (d+1) \over 2 \Lambda}~ {N^{d - 1}
\over P(N)} d N^2 ~+~{ 2 \Lambda \over d (d + 1)} ~{P(N) \over N^{d
-1}}~{d \theta^2 \over \beta^2}~,
\eeq
where $N$ now varies between the two roots, $N_1$ and $N_2$ of
$P(N)$. The region of variation of the angle $\theta$ depends on
the precise form of $V(z)$ and the choice of coefficient $\beta$;
$\beta$ in (\ref{ttheta}) is an arbitrary constant that can be
reabsorbed in redefinition of $\theta$.\footnote{In  $d=1$, $\alpha
= 0$, $P(N) = N^2 + b$, so upon changing variables $N
= |b|^{1/2} \cos \Omega$  in (\ref{metric2}) we recover the solution
obtained in \cite{DJ} by explicitly solving the equations of
motion.}

Clearly the metric (\ref{metric2}) has singularities
 at $N = N_1,N_2$. To find what these singularities imply
 for the physical
interpretation of the solution, it is convenient to consider again
the metric in the form (\ref{metric1}), (\ref{exp2K}). Clearly the
zeros of $V(z)$ are singular points of the metric. Under the map $z
\rightarrow t$ (\ref{ttheta}) these zeros are mapped to $t = \pm
\infty$, which, finally, are mapped to the roots of the polynomial
$P(N)$, $N_{1,2}$.

Now recall that  the geodesic equation---the equation of motion of
the branes (\ref{geodesic}) has also to be obeyed (for example, if
this equation is not obeyed, an initially static brane will
accelerate). In the static gauge, $X^\mu = \sigma^\mu, X^i_{,\mu} =
0$, (\ref{geodesic})
 implies that $\sqrt{\tilde{g}}
\Gamma^i_{\mu
\nu} \tilde{g}^{\mu \nu} = 0$ at the positions of the branes
(note that in the static gauge the induced metric is equal to the
the metric of the embedding space restricted to the brane).
 For our
ansatz (\ref{metricansatz}) the geodesic equation in the static
gauge amounts to the condition
\beq
\label{condition}
e^{(d-2)\lambda - 2K} \lambda_{,i} = 0 ~\rightarrow ~ N^{d-1} V =
0~,
\eeq
 which should hold at the positions of the branes. The positions of
  the branes, therefore, correspond to zeros of the
  holomorphic function $V$ (\ref{V}),
  and hence to $N_{1,2}$ in the coordinates (\ref{metric2}).

In the following, consider the simple ansatz for $V(z)$, $V(z) =
z/c$; note that since $t
= 2 \Lambda (c \log z + c^* \log \bar{z})/(d+1)d$, single
valuedness of the map $t(z, \bar{z})$ requires $c$ to be real. To
see what kind of singularity this is, consider the $\mu,
\nu$ components of the Einstein equations in the static gauge:
\beq
\label{einsteinmunusource}
\sqrt{\hat{g}}~ \left( \hat{R}_{\mu \nu} - {1\over 2} \hat{g}_{\mu \nu} \hat{R} +
\hat{g}_{\mu \nu} \Lambda \right) ~=~ {1\over 2} {f_a^d\over M^d} ~
\delta^2(y - X_a)~ \tilde{g}^{1/2} ~\tilde{g}_{\mu\nu} ~,
\eeq
where the hats indicate that the curvatures and metric are those of
eq.~(\ref{metricansatz}). The metric in the $z$-coordinates
(\ref{metric1}) can be written in the form:
\beq
\label{metricnearzero1}
d s^2 ~=~ N^2~ d s_{(d)}^2 +  e^{2
\kappa} ~d z ~d {\bar z}~,
\eeq
where
\beq
\label{kappa}
\kappa ~=~ {1 \over 2} \log \left[ {c^2 \over |z|^2} {P(N) \over N^{d-1}} {2 \Lambda \over d
(d + 1) }\right]
\eeq
Near the singularity, equation (\ref{einsteinmunusource})
becomes:
\beq
\label{singul}
- {1\over 2}~   e^{2 \kappa}~ R_{2} ~= - \partial {\bar\partial} \kappa ~=
~ {1\over 2} ~{f_a^d \over M^d}~ \delta^2 (z - X_a) ~,
\eeq
where the equality holds up to terms that do not contain delta
functions. To calculate the function $\kappa$ (\ref{kappa}) near $z
= 0,
\infty$ note that these two points are mapped by
(\ref{ttheta}) to  $t = \pm \infty$ (whether $z=0$ is mapped to $t
= +\infty$ or $-\infty$ depends on the signs of $\Lambda$ and $c$).
These points, on the other hand, correspond to the roots of the
polynomial $P$, $N_1$ and $N_2$. Near $z = 0$ (assume, for
concreteness, that $z = 0$ is mapped to $N
= N_1$), the dependence of $N$ on $t$ is easy to find upon
solving eqn.~(\ref{Finaleqn}) by keeping the most singular term:
\beq
\label{NnearN1}
| N(z) - N_1 | \simeq |z|^{2 \delta_1}~,
\eeq
where
\beq
\label{delta1}
\delta_1 = {2 \Lambda c \over d (d+1)} {P^\prime(N_1) \over N_1^{d-1}}~.
\eeq
Therefore, near $z=0$, the function $\kappa$ from the metric
(\ref{metricnearzero1}) behaves as:
\beq
\label{kappanearzero}
\kappa ~\simeq ~ -  \log |z|^{1 - \delta_1} ~.
\eeq
Upon comparison with eq.~(\ref{singul}), noting that $\partial
\bar\partial \kappa  = - 2 \pi (1 - \delta_1) \delta^2 (z)$
we can now find the tension of the brane at $z = 0$,
\beq
\label{tensionnearN1}
f_1^d = M^d ~4  \pi ~(1 - \delta_1) ~.
\eeq

We can repeat the same analysis near $z = \infty$ mapped to the
other root, $N=N_2$.\footnote{The leftmost equation in
(\ref{condition}) implies that the roots of $P(N)$ are stationary
points of $\lambda$ and the equation of motion of the branes is
also satisfied at $N_2$.} It is convenient to change variables to
$u = 1/z$. Note that the metric (\ref{metricnearzero1}) (and the
function $\kappa$) has the same form in the $u$ variables. The
behavior of $N$ near $u
= 0$ from eq.~(\ref{Finaleqn}) is:
\beq
\label{NnearN2}
|N(u) - N_2 |  \simeq  |u|^{- 2 \delta_2}~,
\eeq
where
\beq
\label{delta2}
\delta_2 = {2 \Lambda c \over d (d+1)} {P^\prime(N_2) \over N_2^{d-1}}~.
\eeq
Consequently, $\kappa \simeq -
\log |u|^{1 + \delta_2}$, and the tension of the brane near $z =
\infty$ is then
\beq
\label{tensionnearN2}
f_2^d = M^d 4  \pi ~(1 + \delta_2) ~.
\eeq

We note the conditions that $0 < \delta_1 < 1$ and $-1 < \delta_2 <
0$. They follow from requiring positivity of the brane tensions
($\delta_2>-1$ and $\delta_1 <1 $).  On the other hand, the
distance between the two branes should be finite along any path in
 the  $z$-plane, which requires $\delta_1  > 0$ and
  $\delta_2   < 0$; this follows from considering
 the interval (\ref{metric1}) in the $z$-coordinates and requiring integrability
 of $ds$ near $z = 0$
 and $z = \infty$.\footnote{Equivalently,  $\delta_1 > 0$,
 $\delta_2 < 0$ ensure that the map $|z| \rightarrow N$ is
 nondegenerate near the ends of the interval, as follows
 from the expressions (\ref{NnearN1}, \ref{NnearN2}).}
Finally, positivity of the metric (\ref{metric1}) of the transverse
space also requires that $\Lambda P(N) N^{1-d}$ be positive as $N$
varies over the interval $N_1, N_2$. These restrictions impose
constraints on the various parameters in the action and the
solution; these will be considered for our $d = 4$ example in
Section 4.

We can now deduce a general formula for the Planck scale and
cosmological constant on the $a$-th brane. The Einstein term in
d-dimensional effective action will be:
\beqa
\label{Mplanck}
&&- M^{d} ~\int d^d x ~d N ~{d \theta \over \beta} ~ \sqrt{-
\det ( N^2 g_{(d)}) } ~ R_{(d)} ( N^2 g_{d}) \\
 ~&=&
- \int d^d x \sqrt{g_{(d)}}~ R_{(d)} (g_{d}) ~ M^{d}
~ \int_{N_1}^{N_2} N^{d-2} d N  ~\beta^{-1} \int d \theta
~\nonumber ~.
\eeqa
Let   the warp factor on  the $a$-th brane be $N_{(a)}$. The
physical metric there is the induced metric $N_{(a)}^2 g_{(d)}$.
Hence the d-dimensional Planck constant---the coefficient in front
of the curvature term in the action---will be:
\beq
\label{dplanck}
M_{Pl(a)}^{d-2} ~=~{M^{d} \over (d-1)} ~N_{(a)}^{2-d}~\left(
N_2^{d-1}
- N_1^{d-1} \right) ~ \beta^{-1} \int d \theta   ~,
\eeq
where the $\theta$ integral is over the appropriate range that
follows from (\ref{ttheta}) and depends on the choice of $V$ and
$\beta$. Since the induced metric on the $a$-th brane is $N_{(a)}^2
g_{(d)}$, and since $g_{(d)}$ obeys the equation (\ref{4dmetric}),
the cosmological constant on the $a$-th brane, denoted by
$\Lambda_{(d),(a)}$,   is
\beq
\label{lambdad}
\Lambda_{(d),(a)} ~=~ N_{(a)}^{-2}~ \alpha~.
\eeq
Similarly, one can derive an expression for the proper distance
between the branes at $N_1$ and $N_2$, which we will present for
the case $d=4$   in the following section.

\section{An explicit example and some speculations.}

As an application of the general analysis of the previous section,
let us now consider the physically interesting case $d = 4$. The
polynomial $P(N)$ then is of degree 5. It is easy to see that if
the coefficients $a, b$ in $P(N)$ obey\footnote{More precisely, the
numerical constant on the r.h.s. is $(3/5)^{3/2} - (3/5)^{5/2}$.}
\beq
\label{abconditions}
a > 0 ~~ {\rm and} ~~ 0 < b < .18 ~ a^{5/2}  ~
\eeq
then there are two positive roots $N_1$ and $N_2$ of $P(N)$; in
addition, $P(N) < 0$ as $N$ varies between the two roots, while
$P^\prime (N_1) < 0 $ and $P^\prime (N_2) > 0$.
 Since $P(N)$ is negative in the range of variation of
$N$, the metric (\ref{metric1}) is positive definite only for
negative $\Lambda$,
 corresponding to de Sitter space in our convention.\footnote{This conclusion
 is always true: in the case of two negative roots of $P(N)$, which can
 be achieved by adjusting the value of $b$, $P>0$ between the roots, but
 $N < 0$ and positivity of the metric also requires negative $\Lambda$.}
 Note that since
$a \sim
\alpha/\Lambda > 0$, this implies that $\alpha < 0$ as well and
 the four dimensional metric
 on the branes is also de Sitter (also note that a solution with
$\alpha = 0$ does not exist---then  $P(N)$  has a single real root
and the equation (\ref{Finaleqn}) can not be solved for the desired
interval of values of $t$).

We need to satisfy the conditions $\delta_i > 0$ for the root $N_i$
 which is the image of $z = 0$ and $\delta_j < 0$ for the
other root $N_j$ (the image of $z =\infty$), so that the proper
distance between the two roots is finite. Since $\delta_i
\sim \Lambda c P^\prime (N_i)/N_i^3$, (\ref{delta1}, \ref{delta2}),
 for our choice of two positive
roots, $0< N_1 < N_2$, we find that ${\rm sign}~ \delta_1 = {\rm
sign}~ c$ and ${\rm sign} ~\delta_2 = - {\rm sign}~ c$ (since
$\Lambda < 0$). On the other hand, since $t \sim \Lambda c
\log|z|$, if $c > 0$, then $z  = 0$ is mapped to $t = + \infty$
(and, since $f(N) < 0$, the smaller root $N_1$), while $z = \infty$
is mapped
 to $t = - \infty$ and, hence,  the larger root $N_1$.
 Then the
finite proper distance requirement is satisfied by the choice
$c>0$, yielding  $\delta_1 > 0$, $\delta_2 < 0$.
 The coordinate $\theta$ (\ref{ttheta}) changes then from $(0, 2
\pi c)$ (we chose $\beta = - 1/2$). Expressing the parameters
$\delta_{1,2}$ through the tensions of the branes we obtain:
\beqa
\label{tensions4d}
    \delta_1 &=&  {|\Lambda| c \over 10}
    ~ { |5 N_1^4 - 3 a N_1^2| \over N_1^3} = 1 - {f_1^4 \over 4 \pi M^4} \nonumber \\
     |\delta_2| &=& {|\Lambda| c \over 10}
     ~ { 5 N_2^4 - 3 a N_2^2 \over N_2^3}= 1 - {f_2^4 \over 4 \pi M^4}~.
\eeqa
The induced metric on the $i$-th brane has
\beq
\Lambda_{4,i} =
N_{i}^{-2} \alpha ~~{\rm and}~~ M_{Pl, i}^2 = M^4~ {4 \pi c \over
3} ~ N_i^{-2} ~ (N_2^3 - N_1^3)~,
\eeq
while the proper distance between the branes is:
\beq
\label{properdistance}
R ~=~ \sqrt{10 \over \Lambda} ~ \int_{N_1}^{N_2}  { N^{3/2} d N
\over | N^5 - a N^3 + b |^{1/2}}~.
\eeq

There are four parameters in the action with which we started: the
6 dimensional gravitational constant, $M$, the cosmological
constant, $\Lambda$, and the tensions of the two branes, $f_1,
f_2$. The solution involves three additional dimensionful
parameters, the parameter $\alpha$ of the ansatz for the
d-dimensional metric (\ref{4dmetric}), the integration constant
$\epsilon$ (\ref{finaleqn2}), and $c$---the ``strength" of the zero
of $V(z)$. The equations (\ref{tensions4d}) expressing the tensions
for the branes through the other parameters of the solutions
represent two constraints on the seven parameters $M, \Lambda, f_1,
f_2, \epsilon, \alpha, c$. We can use them to eliminate the
tensions $f_{1,2}$ from the list of our parameters. The only
importance of eqns.~(\ref{tensions4d}) then is that their left hand
sides have to be smaller than 1, to ensure positivity of the brane
tensions.

{}From the four dimensional effective theory point of view the
quantities of interest are the four dimensional Planck constant,
$M_{Pl}$ and the  cosmological constant, $\Lambda_{4}$. Of interest
are also the mass of  Kaluza-Klein excitations as well as the
distance scale where a gravitational experiment on the brane will
reveal deviations from Newton's law. The last two quantities
depend, of course on the size of the extra dimensions (and hence on
the distance $R$ (\ref{properdistance})). However, due to the
  warped geometry, there can be also nontrivial dependence on the
warp factors;  we leave a detailed investigation of this issue for
future work.

To get an idea as to what the effect of the nontrivial warp factors
might be, consider first the case where the parameters $a,b$ in
$P(N)$ (\ref{polynomial}), (\ref{ab}) satisfy $0 < b << a^{5/2}$.
The advantage of this limit is that the roots can be easily
 approximated by:
\beq
\label{roots4d}
 N_1 \simeq \left( b\over a
\right)^{1/3} \sim \left( \epsilon \over \alpha
\right)^{1/3} ~~~{\rm and} ~~ N_2 \simeq a^{1/2} \sim \left( \alpha \over \Lambda \right)^{1/2}~.
\eeq
Note that  the case we are considering corresponds to $N_2 \gg
N_1$---the warp factors on the two branes can be (upon adjusting
the parameters, of course) very different.

 The conditions of positive tensions (\ref{tensions4d})
then imply
\beq
\label{positivetensions4d}
{3 |\Lambda| c \over 10 }~  a ~\left( a \over b \right)^{1/3} < 1
~~{\rm and} ~~ |\Lambda| ~c  ~a^{1/2} < 1~.
\eeq
In the case we are discussing, the proper
 distance (\ref{properdistance}) between the branes is easily seen
to be approximately
\beq
\label{properdistance4d}
R \sim {1 \over \sqrt{\Lambda}}~.
\eeq
The induced metrics on the two branes at $N_1$ and $N_2$ are de
Sitter with cosmological constants:
\beq
\label{cc4d}
   ~ |\Lambda_{(4),1}| = N_1^{-2} |\alpha| \sim |\alpha | ~  \left( \alpha \over \epsilon \right)^{2/3}~ ~{\rm and}~~
   |\Lambda_{(4), 2}| = N_2^{-2} |\alpha| \sim |\Lambda|~ ,
\eeq
while the corresponding Planck constants are:
\beq
\label{mpl4d}
M_{Pl,1}^2 \sim  M^4  c ~ a^{3/2} N_1^{-2} \sim M^4 c~
{|\alpha|^{13/6}
\over |\Lambda|^{3/2} |\epsilon|^{2/3} }~
~{\rm and} ~~ M_{Pl, 2}^2
\sim M^4  c ~ a^{3/2} N_2^{-2} \sim M^4 c~ {|\alpha|^{1/2} \over |\Lambda|^{1/2}} ~.
\eeq

The vacuum energy density on the brane is of order $M_{Pl}^2
\Lambda_{(4)}$. If one of the branes described the physical universe,
 the parameters should be tuned so that
 $\Lambda_{(4),i}/M_{Pl,i}^2 \sim 10^{-120}$ (or to a value smaller than
that; note also that this ratio is the same on both branes, since
the warp factors cancel between numerator and denominator). It is
not our objective to solve the cosmological constant problem; we
will only ask whether our parameter space allows for at least a
fine-tuned  situation, subject to the constraints imposed by the
existence of a solution.

Tuning the cosmological constant to the observed small value,
therefore, requires that
\beq
\label{cctuningcondition}
{M^4 \over \Lambda^2}~ (|\Lambda| c) ~a^{1/2}
\sim 10^{120}~.
\eeq
It is convenient to work in terms of the dimensionless ratios $a$,
$b$, $(M^2 |\Lambda|^{-1})$, $(|\Lambda| c)$, and keep $\alpha$ as
the only dimensionful parameter.

Consider first the case where the observed universe is at the first
brane. Then,
$$
M_{Pl, 1}^2
\sim \left[ {M^4\over|\Lambda|^2}~ (|\Lambda| c)~ a^{1/2} \right] ~|\alpha| ~ \left(a\over b\right)^{2/3}
\sim 10^{120} ~|\alpha| ~\left(a\over b\right)^{2/3} ~\sim 10^{54} ~~eV^2~.
$$
Hence,
\beq
\label{alpha3}
\alpha \sim \left( b\over a \right)^{2/3} 10^{- 66} eV^2~.
\eeq
Then
$$
{1\over R}  \sim |\Lambda|^{1/2} \sim |\alpha|^{1/2} a^{-1/2} \sim
{ b^{1/3}\over a^{5/6}} ~10^{-33} ~ eV~.
$$
This is a very large distance, unless $b\gg a^{5/2}$, which violates
(\ref{abconditions}). If the observed universe is on the other
brane, we obtain instead $|\alpha| \sim a 10^{-66} eV^2$ and $1/R
\sim 10^{-33} eV$---also a large value for $R$.

Therefore, if the deviations from Newton's law reveal themselves to
an experimentalist on the brane at a distance scale of order $R$,
then we can not accommodate the observed universe on one of our
branes. If, on the other hand, the relevant scale of deviations is
$N_i R$, then it is possible to fine-tune the parameters to obtain
a value for $N_i R$ of order a millimeter, and hence accommodate the
observed world in our setup (admittedly with enormous fine-tuning
of parameters!).

The above considerations apply also in a more general situation,
where eq.~(\ref{cctuningcondition}) is replaced by:
\beq
\label{cctuninggeneral}
{M^4 \over \Lambda^2}~ (|\Lambda| c) ~{N_2^3 - N_1^3 \over a}
\sim 10^{120}~,
\eeq
while (\ref{alpha3}) becomes:
\beq
\label{alpha4}
|\alpha| ~\sim~ N_i^2~ 10^{- 66} eV^2~,
\eeq
provided that the observed universe is identified with the brane at $N_i$. But
then the inverse radius becomes:
\beq
\label{radius}
{1 \over R} ~\sim~ {N_i \over a^{1/2}} ~I~ 10^{-33} eV~,
\eeq
where $I$ is the integral
\beq
\label{integral}
I ~=~
\int_{N_1}^{N_2}  { N^{3/2}
d N
\over | N^5 - a N^3 + b |^{1/2}}~,
\eeq
which is trivial in the limit $b \rightarrow 0$ considered before.

Assuming that in our geometrical setup the deviations of gravity
from Newton's law will show up, to an observer on ``our" brane, at
a distance scale of order $R$, an acceptable value of $1/R
> 10^{-3} eV$ requires $
N_i I a^{-1/2} > 10^{30}$ (which appears hard to achieve, similar
to the example given above). If the deviations show up at a scale
$N_i R$, then the requirement is $I a^{-1/2} > 10^{30}$ (which is
achievable via fine tuning). We leave the
investigation of this issue for future work.

\section{Acknowledgments.}

We would like to thank N. Kaloper, R. Leigh, M. Luty, and V.
Moncrief for helpful discussions and comments.

\end{document}